\documentclass[prb,twocolumn,superscriptaddress,showpacs,amsmath,amssymb]{revtex4}

\usepackage{graphicx} 
\usepackage{dcolumn} 

\def\bG{{\bf G}}

\def\bk{{\bf k}}

\def\bp{{\bf p}}
\def\bq{{\bf q}}

\def\bG{{\bf G}}
\def\bQ{{\bf Q}}

\def\b0{{\bf 0}}
\def\btau{{\boldsymbol\tau}}

\def\Re{{\rm Re}}
\def\Im{{\rm Im}}

\def\alf{\alpha}
\def\eps{\epsilon}
\def\gam{\gamma}
\def\Gam{\Gamma}
\def\lam{\lambda}
\def\Lam{\Lambda}
\def\om{\omega}

\def\sg{\sigma}
\def\Sg{\Sigma}

\def\sgn{{\rm sgn}}

\begin{document}

\title{Fluctuation effects at the onset of $\bf 2k_F$ density wave order \\
with one pair of hot spots in two-dimensional metals}

\author{J\'achym S\'ykora}
\affiliation{Max Planck Institute for Solid State Research,
 D-70569 Stuttgart, Germany}
\author{Tobias Holder}
\affiliation{Department of Condensed Matter Physics, Weizmann Institute of Science, Rehovot, Israel 76100}
\author{Walter Metzner}
\affiliation{Max Planck Institute for Solid State Research,
 D-70569 Stuttgart, Germany}

\date{\today}

\begin{abstract}
We analyze quantum fluctuation effects at the onset of charge or spin density wave order in two-dimensional metals with an incommensurate $2k_F$ wave vector connecting a single pair of hot spots on the Fermi surface. We compute the momentum and frequency dependence of the fermion self-energy near the hot spots to leading order in a fluctuation expansion (one loop). Non-Fermi liquid behavior with anomalous frequency scaling and a vanishing quasi particle weight is obtained. The momentum dependence yields a divergent renormalization of the Fermi velocity and a flattening of the Fermi surface near the hot spots. Going beyond the leading order calculation we find that the one-loop result is not self-consistent. We show that any momentum-independent self-energy with a non-Fermi liquid frequency exponent wipes out the peak of the polarization function at the $2k_F$ wave vector, and thus destroys the mechanism favoring $2k_F$ density waves over those with generic wave vectors. However, a $2k_F$ density wave quantum critical point might survive in presence of a sufficiently flat renormalized Fermi surface.
\end{abstract}
\pacs{71.10.Hf,75.30.Fv,64.70.Tg}

\maketitle


\section{Introduction}

Quantum phase transitions are zero temperature phase transitions which can be tuned by a non-thermal control parameter such as pressure or chemical substitution. In case of continuous transitions, the symmetric and the ordered phase touch at a quantum critical point (QCP). While thermal fluctuations are absent at zero temperature, quantum fluctuations of the order parameter lead to quantum critical phenomena at the QCP. Quantum critical fluctuations also play an important role in a restricted region at low finite temperatures, the so-called quantum critical region near the QCP in the phase diagram.\cite{sachdev99}

Order parameter fluctuations at or near quantum critical points in metals interact strongly with fermionic excitations. They can destroy Landau quasi-particles, thus undermining the basis for Fermi liquid theory.\cite{loehneysen07}
Vice versa, the order parameter fluctuations are themselves strongly affected by the gapless fermionic degrees of freedom in metallic systems. An effective order parameter theory obtained by integrating out the fermions is thus complicated by singular interactions. The assumption of a regular order parameter theory as originally proposed by Hertz \cite{hertz76} and Millis \cite{millis93} is therefore not justified, at least not {\em a priori.} Instead, one is compelled to treat fermions and their order parameter fluctuations on equal footing.

Quantum critical points in metals fall into a variety of distinct universality classes. The geometry of the ordering wave vector $\bQ$ plays a crucial role. 
Magnetic order with $\bQ = \b0$ includes Heisenberg and Ising ferromagnets, while examples for charge order with $\bQ = \b0$ are given by the director nematic in continuum systems \cite{oganesyan01} and the Ising nematic on a square lattice.\cite{metzner03,metlitski10_nem}
Concerning charge and magnetic order with $\bQ \neq \b0$, that is, charge and spin density waves, commensurate and incommensurate wave vectors need to be distinguished, \cite{footnote1} and whether $\bQ$ is a nesting vector of the Fermi surface or not. The most thoroughly analyzed quantum critical point with $\bQ \neq \b0$ is the one at the onset of commensurate antiferromagnetic order with a non-nested wave vector $\bQ = (\pi,\pi)$ in two dimensions.\cite{abanov03,metlitski10_af}
Quantum critical behavior at the onset of charge order with a non-nested incommensurate wave vector has been studied in the context of cuprate superconductors.\cite{castellani95}

A special situation arises when $\bQ$ is a {\em nesting}\/ vector connecting Fermi points with antiparallel Fermi velocities. Spin and charge correlations at such wave vectors feature a well-known singularity generated by an enhanced phase space for low-energy particle-hole excitations. In isotropic systems, the singularity is located at wave vectors $\bQ$ with $|\bQ| = 2k_F$, where $k_F$ is the radius of the Fermi surface. In inversion symmetric crystalline metals with a valence band dispersion $\eps_\bk$, wave vectors $\bQ$ obeying the condition $\eps_{(\bQ+\bG)/2} = \eps_F$ are nesting vectors. Here $\eps_F$ is the Fermi energy, and $\bG$ is a reciprocal lattice vector. We refer to such nesting vectors also as ``$2k_F$ vectors''.
The above notion of nesting should not be confused with the stronger condition of {\em perfect nesting}, where a momentum shift $\bQ$ maps extended Fermi surface pieces on top of each other. Perfect nesting is possible only for special band structures and electron densities.

$2k_F$ singularities are particularly pronounced in metals with reduced dimensionality. Charge and spin susceptibilities in low-dimensional systems frequently exhibit a peak at $2k_F$ vectors, such that these wave vectors are favored for charge- and spin-density wave instabilities. For example, the ground state of the two-dimensional Hubbard model undergoes spin-density wave instabilities at $2k_F$-vectors upon increasing the coupling strength, at least within mean-field (Hartree-Fock) theory.\cite{schulz90,igoshev10}
Also $d$-wave bond charge order generated by antiferromagnetic fluctuations in models for cuprate superconductors \cite{metlitski10_af,sachdev13} occurs naturally at $2k_F$ vectors.\cite{holder12,punk15}
Although weaker in three dimensions, the $2k_F$ singularity in the spin susceptibility was recently found to determine the wave vector and the quantum critical behavior of the dominant spin density wave instability of the three-dimensional Hubbard model with a large coupling strength. \cite{schaefer17}

Quantum critical behavior at the onset of density wave order with a $2k_F$-vector in two-dimensional metals was first analyzed by Altshuler {\em et al.} \cite{altshuler95} They derived non-Fermi liquid power-laws for the fermion self-energy and the susceptibility for the case where $\bQ$ is half a reciprocal lattice vector. In the special case where $\bQ=(\pi,\pi)$ is a $2k_F$ vector, additional umklapp processes have to be taken into account.\cite{bergeron12,wang13}
For incommensurate $2k_F$ vectors, Altshuler {\em et al.}\ found very strong infrared divergences in two-loop contributions to the susceptibility. They therefore concluded that the order-parameter fluctuations destroy the quantum critical point and replace it by a first order transition.

Recently, two of us have analyzed the influence of quantum critical fluctuations at the onset
(QCP) of charge- or spin-density wave order with an incommensurate $2k_F$ wave vector in two-dimensional metals on single-particle excitations.\cite{holder14} To this end, the fermion self-energy was computed to first order (one loop) in the fluctuation propagator, which was also computed in a one-loop approximation. A breakdown of Fermi liquid behavior was obtained at {\em hot spots} of the Fermi surface, that is, Fermi surface points that are connected by the ordering wave vector $\bQ$. Two qualitatively distinct cases had to be distinguished. In the first case, the ordering wave vector $\bQ$ connects only a single pair of hot spots and points typically in axial or diagonal direction (see Fig.~1). The frequency dependence of the one-loop self-energy at the hot spots obeys a power-law with exponent $\frac{2}{3}$ in this case.\cite{holder14} In the second case, $\bQ$ connects two pairs of hot spots, and the imaginary part of the real frequency one-loop self-energy features a linear frequency dependence at the hot spots.\cite{holder14}
\begin{figure}
\centering
\includegraphics[scale=0.833]{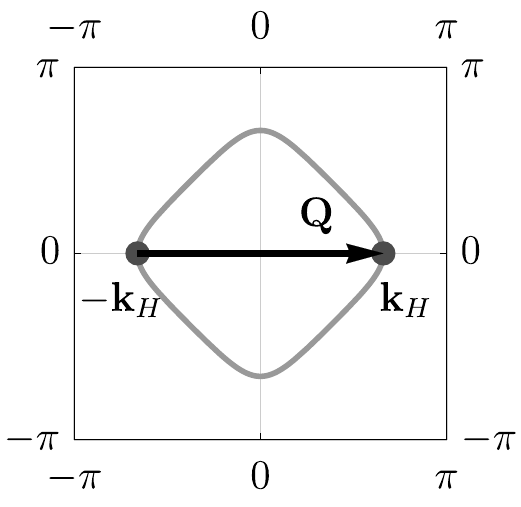}
\caption{Fermi surface with a single pair of hot spots $\pm\bk_H$ connected by ordering wave vector $\bQ$.}
\end{figure}
Note that at one-loop order actually no distinction needs to be made between the case of incommensurate $\bQ$ and the case of commensurate $\bQ$, as long as $\bQ$ is not half a reciprocal lattice vector.

In this paper we extend the analysis of the QCP for the case of a single hot-spot pair in several directions. First, we compute the momentum dependence of the one-loop self-energy at the QCP in the vicinity of the hot spots. In particular, we thereby determine the non-Fermi liquid regime in energy-momentum space near the hot spots. Second, we analyze to what extent the one-loop susceptibility and fluctuation propagator is modified by the singular fermion self-energy. We find that this feedback effect is substantial, giving rise to various possible scenarios.

The article is structured as follows. In Sec.~II we discuss the asymptotic behavior of the RPA susceptibility and effective interaction at the QCP. In Sec.~III the momentum and frequency dependence of the fermion self-energy near the hot spots as obtained to leading order in the effective interaction (one-loop) is presented. In Sec.~IV the issue of self-consistency and stability of the QCP is investigated. Final conclusions follow in Sec.~V. The appendices contain a derivation of effective interactions for two familiar microscopic models (Appendix A), and the evaluation of the loop integrals (Appendixes B and~C).


\section{RPA susceptibility and effective interaction}

We consider a one-band system of interacting fermions with a bare dispersion relation $\eps_\bk$. Our calculations are based on the standard quantum many-body formalism with an imaginary frequency representation of dynamical quantities.\cite{negele87}
The bare fermion propagator is thus given by
\begin{equation}
 G_0(\bk,ik_0) = \frac{1}{ik_0 - \xi_\bk} \\ ,
\end{equation}
where $k_0$ is the frequency variable, and $\xi_\bk = \eps_\bk - \mu$ is the single-particle energy relative to the chemical potential.

We assume that, in mean-field theory, the system undergoes a charge or spin density-wave instability with an incommensurate and nested ($2k_F$) modulation vector $\bQ$ at a QCP which can be reached by tuning a suitable parameter such as density or interaction strength.
Approaching the QCP from the normal metallic regime, the instability is signalled by a diverging RPA susceptibility
\begin{equation} \label{chi}
 \chi(\bq,iq_0) = \frac{\chi_0(\bq,iq_0)}{1 + g \chi_0(\bq,iq_0)} \, ,
\end{equation}
where $g < 0$ is the coupling constant parametrizing the bare interaction in the instability channel, and $\chi_0$ is the bare susceptibility
\begin{eqnarray} \label{chi_0}
 \chi_0(\bq,iq_0) &=& - N \, \Pi_0(\bq,iq_0) \nonumber \\
 &=& - N \int \frac{d^2\bk}{(2\pi)^2} \int \frac{dk_0}{2\pi} \, f_{\bk-\bq/2}^2
 \nonumber \\
 &\times& G_0(\bk,ik_0) \, G_0(\bk-\bq,ik_0-iq_0) \, .
\end{eqnarray}
$N$ is the spin-multiplicity ($N=2$ for spin-$\frac{1}{2}$ fermions), and $f_\bp$ is a form factor related to the internal structure of the density-wave order parameter. For an order parameter with $s$-wave symmetry, $f_\bp$ is symmetric under rotations and reflections. In the following we assume $f_\bp = 1$ for definiteness. A generalization to form factors with other symmetries such as $d$-wave is straightforward.
Note that Eq.~(\ref{chi}) holds also for the spin susceptibility in the normal (symmetric) phase of spin-rotation invariant systems, where all components of the spin susceptibility are equal. For the Hubbard model, the RPA spin susceptibility is given by Eq.~(\ref{chi}) with $g=-U/2$.

\begin{figure}
\centering
\includegraphics[scale=0.833]{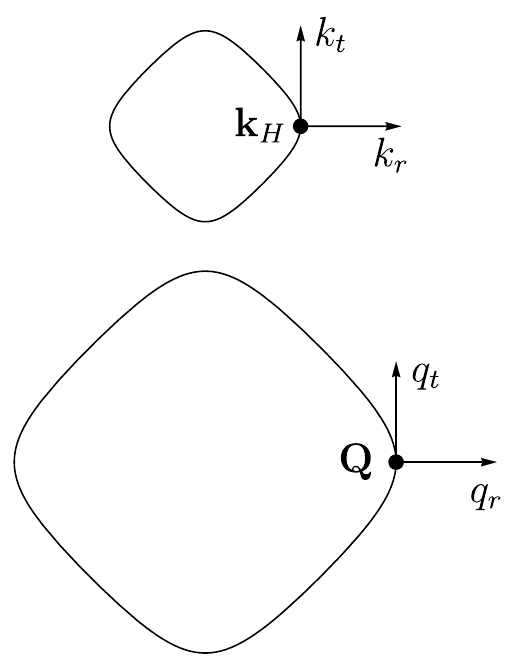}
\caption{(Top) Normal and tangential momentum coordinates for momenta near one of the hot spots on the Fermi surface. (Bottom) Normal and tangential momentum coordinates for momenta near the ordering vector $\bQ$ on the $2k_F$-line (extended-zone scheme without backfolding into the first Brillouin zone).}
\end{figure}
Throughout this article we consider the case where the ordering wave vector $\bQ$ connects a single pair of hot spots $\bk_H$ and $-\bk_H$ on the Fermi surface.
To parametrize momenta near the ordering momentum $\bQ$ we introduce coordinates normal and tangential to the line formed by the $2k_F$-vectors in $\bq$-space, which we denote by $q_r$ and $q_t$, respectively (see Fig.~2).
For momenta near the ordering momentum $\bQ$ and low frequencies, the bare susceptibility can then be expanded as \cite{altshuler95,holder14}
\begin{equation} \label{chi_0_exp}
 \chi_0(\bq,iq_0) =
 \chi_0(\bQ,0) - a \, h(e_\bq,q_0) + b \, e_\bq - c \, \frac{q_t^2}{4m} \, ,
\end{equation}
where $a$, $b$, and $c$ are positive constants, and
\begin{eqnarray}
 h(\bq,q_0) &=& \sqrt{e_\bq + iq_0} + \sqrt{e_\bq - iq_0} \nonumber \\
 &=& \sqrt{2} \sqrt{\sqrt{e_\bq^2 + q_0^2} + e_\bq} \, .
\end{eqnarray}
The energy-momentum relation $e_\bq$ is given by
\begin{equation}
e_\bq = v_F q_r + \frac{q_t^2}{4m} \, ,
\end{equation}
where $v_F$ is the Fermi velocity at $\pm\bk_H$ and $m$ parametrizes the Fermi surface curvature at these points ($mv_F$ is the radius of curvature);
$e_\bq/v_F$ is the oriented distance of $\bq$ from the $2k_F$-line.
In previous works \cite{altshuler95,holder14} the last term in the expansion (\ref{chi_0_exp}) was discarded. It plays however a significant role as it defines the maximum of $\chi_0(\bq,0)$ at $\bQ$ with respect to variations along the $2k_F$-line.
The prefactor of the square-root term is determined by the Fermi velocity and curvature near $\pm \bk_H$ as $a = N \sqrt{m}/(4\pi v_F)$, while the other constants $b$ and $c$ receive contributions from everywhere. For fermions with a parabolic dispersion in the continuum and a constant form factor, $b$ and $c$ vanish.\cite{stern67}
At the QCP one has $g \chi_0(\bQ,0) = - 1$, so that the RPA susceptibility assumes the singular form
\begin{equation}
 \chi(\bq,iq_0) = - \frac{g^{-1} \chi_0(\bQ,0)}
 {a \, h(e_\bq,q_0) - b \, e_\bq + c \, \frac{q_t^2}{4m}} \, . 
\end{equation}

To deal with the critical order parameter fluctuations, the perturbation expansion has to be organized in powers of a dynamical effective interaction. This arises naturally as a boson propagator by decoupling the bare interaction in the instability channel via a Hubbard-Stratonovich transformation.\cite{hertz76,millis93} Alternatively it can be obtained by an RPA resummation of particle-hole bubbles or ladders.
In the simplest case of a charge-density wave instability in a spinless fermion system, the RPA effective interaction can be written as
\begin{equation} \label{D_rpa}
 D(\bq,iq_0) = \frac{g}{1 + g \chi_0(\bq,iq_0)} \, .
\end{equation}
As an example, we derive the effective interaction for the specific case of spinless lattice fermions with a nearest-neighbor interaction in Appendix A.
For a charge-density wave instability in a spin-$\frac{1}{2}$ fermion system, the effective interaction has the diagonal spin structure
\begin{equation}
 D_{\sg'_1\sg'_2\sg_1\sg_2}(\bq,iq_0) =
 \delta_{\sg_1\sg'_1} \delta_{\sg_2\sg'_2} D(\bq,iq_0) \, ,
\end{equation}
where $\sg_1,\sg_2$ ($\sg'_1,\sg'_2$) are the spin indices of the ingoing (outgoing) fermions.
For a spin-density wave, the effective interaction acquires a non-diagonal spin structure. In a spin-rotation invariant system of spin-$\frac{1}{2}$ fermions, it can be written as
\begin{equation} \label{D_spin}
 D_{\sg'_1\sg'_2\sg_1\sg_2}(\bq,iq_0) =
 \btau_{\sg_1\sg'_1} \cdot \btau_{\sg_2\sg'_2} \, D(\bq,iq_0) \, ,
\end{equation}
where $\btau = (\tau^x,\tau^y,\tau^z)$ is the vector formed by the three Pauli matrices $\tau^x,\tau^y,\tau^z$.
We derive the effective interaction for the spin density wave instability in the Hubbard model as an example in Appendix A. 

Expanding the bare susceptibility as in Eq.~(\ref{chi_0_exp}), the effective interaction at the QCP assumes the asymptotic form
\begin{equation}
 D(\bq,iq_0) =
 - \frac{1}{a \, h(e_\bq,q_0) - b \, e_\bq + c \, \frac{q_t^2}{4m}} \, .
\end{equation}
It thus features the same singularity as the RPA susceptibility.
In case that the coupling $g$ has a (regular) momentum dependence for $\bq$ near $\bQ$, the coefficients $b$ and $c$ are not determined by $\chi_0(\bq,0)$ only, but receive additional contributions from the expansion of $g^{-1}(\bq)$ around $\bQ$.


\section{Fermion self-energy}

To leading order in the effective interaction, the fermion self-energy is given by the one-loop expression
\begin{eqnarray} \label{Sg_oneloop}
 \Sg(\bk,i\om) &=& - M \int \frac{d^2\bq}{(2\pi)^2} \int \frac{dq_0}{2\pi} \,
 D(\bq,iq_0) \nonumber \\ 
 &\times& G_0(\bk-\bq,i\om-iq_0) \, ,
\end{eqnarray}
with $M=1$ for a charge density and $M=3$ for a spin density instability.
We evaluate $\Sg(\bk,i\om)$ for low frequencies $\om$ and momenta $\bk$ near a hot spot $\bk_H$.
The dominant contributions come from momentum transfers $\bq$ near $\bQ$, such that $\bk-\bq$ is situated near the antipodal hot spot $-\bk_H$.
We assume that the Fermi surface is convex at $\pm \bk_H$. Using normal and tangential coordinates for $\bk$ near $\bk_H$ as shown in Fig.~2, we expand the dispersion relation to leading order as $\xi_{\bk-\bq} = - v_F (k_r-q_r) + \frac{1}{2m}(k_t-q_t)^2$.
It is convenient to perform the momentum integration by using $q_t$ and $e_\bq$ as integration variables.


\subsection{Imaginary part}

In Appendix B we show that the imaginary part of the one-loop self-energy has the following asymptotic form for low frequencies and momenta near $\bk_H$:
\begin{equation} \label{ImSg}
 \Im\Sg(\bk,i\om) =
 - \frac{M}{\pi N} \, \sgn(\om) \left| \frac{\om}{\bar b} \right|^{2/3}
 I(\tilde\xi_\bk,\tilde k_t) \, ,
\end{equation}
with $\bar b = b/a$ and the dimensionless scaling variables
\begin{equation}
 \tilde\xi_\bk = \frac{\xi_\bk}{(|\om|/\bar b)^{2/3}} \, , \quad
 \tilde k_t = \frac{k_t}{\sqrt{m} (|\om|/\bar b)^{1/3}} \, .
\end{equation}
The dimensionless scaling function $I$ is given by
\begin{align} \label{I}
 I(\tilde\xi_\bk,\tilde k_t) =&
 \sum_{s = \pm 1} \int_{-\infty}^{\min(0,\tilde\xi_\bk)}
 \frac{d\tilde e_\bq}{\sqrt{1 - \tilde\xi_\bk/\tilde e_\bq}} \nonumber \\
 \times & \ln \left[ 1 +
 \frac{|\tilde e_\bq|^{-3/2}}
 {1 + \gam \Big( s \, \tilde k_t/\sqrt{|\tilde e_\bq|} +
 \sqrt{1 - \tilde\xi_\bk/\tilde e_\bq} \, \Big)^2} \right] ,
 \nonumber \\
\end{align}
where $\gam = c/b$.

The integral in Eq.~(\ref{I}) can be evaluated analytically for large and small $\tilde\xi_\bk$ and $\tilde k_t$ (see Appendix B). Analytic results for the self energy can thus be derived in various limiting cases.
In particular, at the hot spot $\bk_H$ we obtain
\begin{equation} \label{ImSg_kH}
 \Im\Sg(\bk_H,i\om) = - \frac{4M \sgn(\om)}{\sqrt{3} N}
 \left| \frac{\om}{\bar b + \bar c} \right|^{2/3} \, ,
\end{equation}
with $\bar c = c/a$. This result is valid also for $\bk \neq \bk_H$ as long as
$\bar b |\xi_\bk|^{3/2} \ll |\om|$ and $\bar b \frac{|k_t|^3}{m^{3/2}} \ll |\om|$.
Eq.~(\ref{ImSg_kH}) is consistent with the result
$\Im\Sg(\bk_H,\om + i0) = - \frac{2M}{\sqrt{3} N} |\om/\bar b|^{2/3}$
for the real frequency self-energy derived for the special case $\bar c = 0$ in Ref.~\onlinecite{holder14}.

For $k_t=0$ and $\xi_\bk < 0$ with $|\om| \ll \bar b |\xi_\bk|^{3/2}$, we obtain
\begin{equation} \label{ImSg_xik-}
 \Im\Sg(\bk,i\om) = - \frac{2M}{N \sqrt{\bar b} \sqrt{\bar b + \bar c}}
 \frac{\om}{\sqrt{|\xi_\bk|}} \, .
\end{equation}
This result remains valid for $k_t \neq 0$, as long as
$\bar b \frac{|k_t|^3}{m^{3/2}} \ll |\om|$.
For $k_t=0$ and $\xi_\bk > 0$ with $|\om| \ll \min(\bar b,\bar c) \, \xi_\bk^{3/2}$, we find
\begin{equation} \label{ImSg_xik+}
 \Im\Sg(\bk,i\om) = - \frac{4M \mbox{Arcsinh}\sqrt{\bar b/\bar c}}
 {\pi N \sqrt{\bar b} \sqrt{\bar b + \bar c}}
 \frac{\om}{\sqrt{\xi_\bk}} \, ,
\end{equation}
which is also applicable for $k_t \neq 0$ with $\bar b \frac{|k_t|^3}{m^{3/2}} \ll |\om|$.
Note the asymmetry of the prefactors in Eqs.~(\ref{ImSg_xik-}) and (\ref{ImSg_xik+}) for $\xi_\bk < 0$ and $\xi_\bk > 0$, respectively.
The frequency range where Eq.~(\ref{ImSg_xik+}) is valid is restricted not only by $\bar b \, \xi_\bk^{3/2}$ but also by $\bar c \, \xi_\bk^{3/2}$, and thus shrinks to zero for $\bar c = 0$. The above results were derived for a convex Fermi surface. For a concave Fermi surface, Eq.~(\ref{ImSg_xik-}) holds for $\xi_\bk > 0$ and Eq.~(\ref{ImSg_xik+}) for $\xi_\bk < 0$.

Finally, for $\xi_\bk = 0$ and $k_t \neq 0$ with
$|\om| \ll \min(\bar b,\bar c) \frac{|k_t|^3}{m^{3/2}}$, we obtain
\begin{equation} \label{ImSg_kt}
 \Im\Sg(\bk,i\om) = - \frac{2M\sqrt{m}}{N \sqrt{\bar b \bar c}} \,
 \frac{\om}{|k_t|} \, .
\end{equation}
This result is also applicable for $\xi_\bk \neq 0$, as long as $\bar b |\xi_\bk|^{3/2} \ll |\om|$.
Eqs.~(\ref{ImSg_kH}) and (\ref{ImSg_kt}) are also valid if the Fermi surface is concave instead of convex at $\pm\bk_H$.

The results Eqs.~(\ref{ImSg_kH}) and (\ref{ImSg_xik-}) are applicable also for $\bar c=0$, while the limit $\bar c \to 0$ can obviously not be taken in Eqs.~(\ref{ImSg_xik+}) and (\ref{ImSg_kt}). None of these results is applicable for $\bar b = 0$.
In particular, the results do not apply to an isotropic parabolic dispersion and a momentum independent coupling constant, for which $\bar b = \bar c = 0$. In this case the bare static susceptibility is constant for all momenta $\bq$ with modulus $|\bq| \leq 2k_F$. Hence, the RPA propagator diverges for all these momenta at the mean-field QCP, so that an expansion based on the RPA is ill-defined.

At the hot spots the imaginary part of the self-energy vanishes sublinearly for vanishing $\om$. Hence, there are no stable quasi-particle excitations at these points, which implies a breakdown of Fermi liquid theory. At momenta away from the hot spots, $\Im\Sg(\bk,i\om)$ is linear in $\om$ as in a Fermi liquid.\cite{footnote2} We can therefore define a momentum dependent $Z$-factor
\begin{equation}
 Z_\bk = \left( 1 - 
 \left. \frac{\partial\Sg(\bk,i\om)}{i\partial\om} \right|_{\om=0} \right)^{-1} \, .
\end{equation}
In a Fermi liquid, $Z_\bk$ is finite on the Fermi surface ($\bk=\bk_F$) and describes the spectral weight of quasi-particle excitations. For the one-loop self-energy at the $2k_F$ QCP derived above, $Z_\bk$ is well-defined and finite for $\bk \neq \pm \bk_H$, but it vanishes as $|k_t|$ upon approaching the hot spot along the Fermi surface, and as $\sqrt{|\xi_\bk|}$ upon approaching $\bk_H$ perpendicularly to the Fermi surface.


\subsection{Real part}

There is no non-Fermi liquid contribution to the frequency dependence of the real part of the self-energy at the hot spots. The symmetry of 
$\Im\Sg(\bk_H,\om + i0) = - \frac{2M}{\sqrt{3} N} |\om/\bar b|^{2/3}$ under a sign change of $\om$ implies that there is no contribution of order $|\om|^{2/3}$ to
$\Re\Sg(\bk_H,i\om)$.
However, there are singular (real) contributions to the momentum dependence of
$\Sg(\bk,0)$, which renormalize the dispersion and the Fermi surface shape near the hot spots.
We parametrize the momentum dependence by the variables $\xi_\bk$, which describes the distance from the Fermi surface, and $k_t$.
For momenta close to a hot spot $\bk_H$ in {\em perpendicular}\/ direction to the Fermi surface (that is, $k_t=0$), one obtains (see Appendix B)
\begin{eqnarray} \label{ReSg_xik}
 \delta\Sg(\bk,0) &=& \Sg(\bk,0) - \Sg(\bk_H,0) \nonumber \\
 &=& \frac{M}{N} \left[ \frac{1}{1 + \Theta(\xi_\bk)} + K(\gam) \right]
 \xi_\bk \ln\frac{\Lam}{|\xi_\bk|} + {\cal O}(\xi_\bk) ,
 \nonumber \\
\end{eqnarray}
where $\Lam$ is an arbitrary fixed energy scale, $\gam = c/b$, and
\begin{equation}
 K(\gam) =
 \frac{2}{\pi} \left( \frac{\sqrt{\gam}}{1+\gam} -
 \arctan\sqrt{\gam} \right) \, .
\end{equation}
Hence, $\Sg(\bk,0)$ leads to a logarithmically diverging renormalization of the bare dispersion $\xi_\bk$. The prefactor in Eq.~(\ref{ReSg_xik}) is asymmetric, that is, it depends on the sign of $\xi_\bk$.
The function $K(\gam)$ ranges from $0$ for $\gam = 0$ to $-1$ for $\gam \to \infty$, and it crosses the value $-1/2$ at $\gam = \gam_c \approx 5.1$. For $\xi_\bk < 0$, the self-energy makes the effective dispersion steeper, with a logarithmically diverging slope at the hot spots. For $\xi_\bk > 0$, it becomes steeper only as long as $\gam < \gam_c$.
By contrast, for $\gam > \gam_c$ the logarithmic self-energy correction reduces the effective dispersion outside the bare Fermi surface (where $\xi_\bk > 0$), and even leads to a sign change upon approaching the hot spots, such that the Fermi surface at the hot spots is destroyed. However, the logarithmic divergence indicates a breakdown of perturbation theory for momenta close to the hot spots, such that the above results are questionable in the region where the logarithm is large.

Moving away from the hot spot {\em along}\/ the Fermi surface (that is, keeping $\xi_\bk = 0$), the self-energy varies as
\begin{eqnarray} \label{ReSg_kt}
 \delta\Sg(\bk,0) &=& \Sg(\bk,0) - \Sg(\bk_H,0) \nonumber \\
 &=& - \frac{M}{N} \, \frac{8\gam^{3/2}}{\pi(1+\gam)^2} \, \frac{k_t^2}{2m} \,
 \ln \frac{2m\Lam}{k_t^2} \, .
\end{eqnarray}
This result has been extracted from a numerical evaluation of the self-energy integral. We refrained from performing a lengthy analytic derivation.
Again we encounter a logarithmic divergence. Due to the negative prefactor, this self-energy correction shifts the Fermi surface into the region where the bare dispersion is positive, which leads to a flattening of the Fermi surface near the hot spots.

Summing the contributions from Eqs.~(\ref{ReSg_xik}) and (\ref{ReSg_kt}), one obtains the total self-energy correction $\delta\Sg(\bk,0)$ for $k_r \neq 0$ and $k_t \neq 0$ in the form
\begin{equation}
 \delta\Sg(\bk,0) = A_r^{\sgn(\xi_\bk)} v_F k_r \ln\frac{\Lam}{v_F|k_r|} +
 A_t^{\sgn(\xi_\bk)} \frac{k_t^2}{2m} \ln\frac{2m\Lam}{k_t^2} ,
\end{equation}
where
\begin{eqnarray}
 A_r^\alf &=& \frac{M}{N} \left[ \frac{1}{1 + \Theta(\alf)} + K(\gam) \right] \, , \\
 A_t^\alf &=& A_r^\alf - \frac{M}{N} \frac{8\gam^{3/2}}{\pi(1+\gam)^2} \, .
\end{eqnarray}
We have checked numerically that this formula describes the correct leading behavior for small $k_r$ and $k_t$ not only in radial and tangential direction, but along any curve where $k_r \propto k_t^2$ in the momentum plane near the hot spots.
Note that $A_t^\alf$ receives a contribution from Eq.~(\ref{ReSg_xik}) with $\xi_\bk > 0$, since the term in Eq.~(\ref{ReSg_xik}) depends on $\xi_\bk$, not only on $k_r$ (see the derivation in Appendix B).
While $A_r^-$ is always positive, the signs of $A_r^+$, $A_t^+$ and $A_t^-$ depend on $\gam$.
$A_r^+$ is positive for $\gam < \gam_c$ and negative for $\gam > \gam_c$ with $\gam_c \approx 5.1$, as already discussed above. $A_t^+$ is positive for $\gam < \gam'_c$ and negative for $\gam > \gam'_c$ with $\gam'_c \approx 0.51$.
Hence, for $\gam > \gam'_c$ the logarithmic divergence of the prefactor of the quadratic tangential momentum dependence flips the Fermi surface shape from convex to concave in the immediate vicinity of the hot spots. However, this effect is also most likely an artifact of applying a perturbative expression outside its range of validity.

The self-energy correction renormalizes the bare dispersion $\xi_\bk$ to become an effective dispersion, which can be written as
\begin{equation}
 \bar\xi_\bk = \xi_\bk + \delta\Sg(\bk,0) =
 Z_r^{\sgn(\xi_\bk)} v_F k_r + Z_t^{\sgn(\xi_\bk)} \frac{k_t^2}{2m} \, ,
\end{equation}
where
\begin{eqnarray}
 \label{Zr_pert}
 Z_r^\alf &=& 1 + A_r^\alf \ln\frac{\Lam}{v_F|k_r|} \, , \\
 \label{Zt_pert}
 Z_t^\alf &=& 1 + A_t^\alf \ln\frac{2m\Lam}{k_t^2} \, .
\end{eqnarray}

Logarithmic divergences are frequently a perturbative manifestation of power-law behavior. In particular, for $k_t=0$, where $\sgn(\xi_\bk) = \sgn(k_r)$,
Eqs.~(\ref{Zr_pert}) might be just the leading order term corresponding to a power law of the form 
\begin{equation}
 \label{Zr}
 Z_r^\alf = \left( \frac{\Lam}{v_F|k_r|} \right)^{A_r^\alf} \, .
\end{equation}
Expanding Eq.~(\ref{Zr}) in powers of the exponent $A_r^\alf$ one obtains Eq.~(\ref{Zr_pert}).
This {\em ad hoc} resummation can also be derived more systematically from a flow equation for the renormalized Fermi velocity $\bar v_F^\alf = Z_r^\alf v_F$. Inserting the perturbative expression~(\ref{Zr_pert}) and taking a derivative with respect to $|k_r|$ yields
$\frac{\partial \bar v_F^\alf}{\partial |k_r|} = - v_F A_r^\alf \frac{1}{|k_r|}$. Making this flow equation self-consistent by replacing the bare Fermi velocity $v_F$ on the right hand side by the renormalized one, we obtain
\begin{equation}
 \frac{\partial \bar v_F^\alf}{\partial |k_r|} =
 - \bar v_F^\alf A_r^\alf \frac{1}{|k_r|} \, .
\end{equation}
With the renormalization condition $\bar v_F^\alf(k_r) = v_F$ for $v_F|k_r| = \Lam$, this flow equation has the unique solution
\begin{equation}
 \bar v_F^\alf(k_r) = \left( \frac{\Lam}{v_F|k_r|} \right)^{A_r^\alf} v_F \, ,
\end{equation}
where $\alf = \sgn(k_r)$. The renormalized Fermi velocity thus obeys a power-law behavior as a function of $k_r$. For $k_r < 0$ it diverges upon approaching $k_r = 0$. For $k_r > 0$ it also diverges if $\gam < \gam_c$, but it vanishes if $\gam > \gam_c$.

For $k_r=0$ and $k_t \neq 0$ (that is, $\xi_\bk > 0$) we can resum Eq.~(\ref{Zt_pert}) in the same way to obtain a renormalized mass
\begin{equation}
 \frac{1}{2 \bar m(k_t)} =
 \left( \frac{2m\Lam}{k_t^2} \right)^{A_t^+} \frac{1}{2m} \, .
\end{equation}
For $\gam < \gam'_c$, the exponent $A_t^+$ is positive such that $\bar m(k_t)$ vanishes upon approaching the hot spot. For $\gam > \gam'_c$ the exponent is negative and $\bar m(k_t)$ diverges. In contrast to the behavior suggested by the perturbative logarithm there is no sign change of $\bar m (k_t)$ inducing a (rather implausible) change of the Fermi surface shape from convex to concave. Note that $A_t^\alf < A_r^\alf$ for all $\gam > 0$. Hence, the ratio of the renormalizations of tangential and radial momentum dependences always vanishes upon approaching the hot spot, indicating a flattening of the Fermi surface for any $\gam > 0$.


\section{Self-consistency and stability of QCP}

The RPA calculation yields a susceptibility with maxima for certain momenta $\bQ$ on the $2k_F$-line. These momenta determine the modulation of the density wave as obtained in mean-field theory.
In the preceding section we have shown that fluctuations at the QCP give rise to a singular fermion self-energy, which leads to pronounced non-Fermi liquid behavior at the hot spots. Since the square root singularity of the RPA susceptibility and effective interaction at $\bQ$ is determined by states in the hot spot regions, one has to analyze to what extent this singularity is affected by the fermion self-energy.

To this end, we compute the polarization function
\begin{equation} \label{Pi}
 \Pi(\bq,iq_0) = \int \frac{d^2\bk}{(2\pi)^2} \int \frac{dk_0}{2\pi} \,
 G(\bk,ik_0) \, G(\bk-\bq,ik_0-iq_0) \, ,
\end{equation}
with fermion propagators $G$ dressed by the self-energy.
Since contributions with momenta near the hot spots $\pm \bk_H$ are expected to dominate, we first approximate $\Sg(\bk,i\om)$ by its asymptotic low-energy behavior at $\bk_H$, as given by Eq.~(\ref{ImSg_kH}). Under this assumption an analytic evaluation is possible. The influence of the momentum dependence of $\Sg(\bk,i\om)$ will be discussed later.

We actually calculate the polarization function for a generalized form of the self-energy defined as
\begin{equation} \label{Sg_alf}
 \Sg(\bk,i\om) = -i C \sgn(\om) |\om|^\alf \, ,
\end{equation}
where $\frac{2}{3} \leq \alf < 1$ and $C$ is a positive constant. This provides a broader view of the singularity structure and allows us to discuss the fate of the QCP for a possibly renormalized exponent $\alf > \frac{2}{3}$.
The Dyson equation $G^{-1} = G_0^{-1} - \Sg$ yields a dressed propagator given by
\begin{equation}
 G(\bk,i\om) = \frac{1}{iC \sgn(\om) |\om|^\alf - \xi_\bk} \, .
\end{equation}
The linear frequency term stemming from the bare propagator is subleading and has thus been discarded.

In Appendix C we compute the leading momentum dependence of the static polarization function $\delta\Pi(\bq,0) = \Pi(\bq,0) - \Pi(\bQ,0)$ for momenta near $\bQ$. The result is a singular function of $e_\bq$.
For $\frac{2}{3} < \alf < 1$ we obtain
\begin{equation} \label{deltaPi_alf}
 \delta\Pi(\bq,0) = a_{\sgn(e_\bq)} |e_\bq|^{\frac{1}{\alf} - \frac{1}{2}} +
 \mbox{regular terms} \, ,
\end{equation}
where $a_{-} < 0$ and $a_{+} > 0$ are two constants given by Eq.~(\ref{apm}).
For $\alf \to 1$ and $C=1$ we recover the momentum dependence of the bare polarization function
\begin{equation}
 \delta\Pi_0(\bq,0) = \frac{\sqrt{m}}{2\pi v_F} \Theta(e_\bq) \sqrt{|e_\bq|} +
 \mbox{regular terms} \, .
\end{equation}
For $\alf = 2/3$ we find
\begin{equation} \label{deltaPi_2/3}
 \delta\Pi(\bq,0) = \frac{3\sqrt{m}}{32\pi v_F C^{3/2}} e_\bq
 \ln\left| \frac{C\Lam^{2/3}}{e_\bq} \right| + \mbox{regular terms} \, ,
\end{equation}
where $\Lam$ is an ultraviolet cutoff. Inserting the prefactor
$C = \frac{4M}{\sqrt{3} N (\bar b + \bar c)^{2/3}}$ from Eq.~(\ref{ImSg_kH}) yields
\begin{equation} \label{deltaPi_2/3_a}
 \delta\Pi(\bq,0) = \frac{3^{7/4} N^{1/2}}{64 M^{3/2}} (b+c) e_\bq
 \ln\left| \frac{C\Lam^{2/3}}{e_\bq} \right| + \mbox{regular terms} \, .
\end{equation}

The square-root singularity of $\delta\Pi_0(\bq,0)$ for $e_\bq > 0$ is replaced by a power-law with a larger exponent between $\frac{1}{2}$ and $1$ for $\frac{2}{3} < \alf < 1$, and by a linear behavior with a multiplicative logarithmic correction for $\alf = \frac{2}{3}$. The value $\alf=\frac{2}{3}$ obtained for the one-loop fermion self-energy thus defines a marginal case. For $\alf < \frac{2}{3}$ the singular contribution to $\delta\Pi(\bq,0)$ would be super-linear and thus subleading compared to the regular linear contributions.

For $\frac{2}{3} \leq \alf < 1$, the fermion self-energy in the propagators generates a leading singular momentum dependence of the polarization function also for $e_\bq < 0$, with the same functional form as for $e_\bq > 0$. The prefactor of the singular term happens to be {\em negative} for $e_\bq < 0$. Since it dominates over the leading regular contribution of the form $-b e_\bq$, it spoils the minimum of $\Pi(\bq,0)$ at $\bq = \bQ$ and shifts it elsewhere, away from the $2k_F$-line.
Correspondingly, the maximum of the susceptibility is also shifted away from the $2k_F$-line, which is inconsistent with the original assumption on the nature of the QCP.

\begin{figure}
\centering
\includegraphics[scale=0.833]{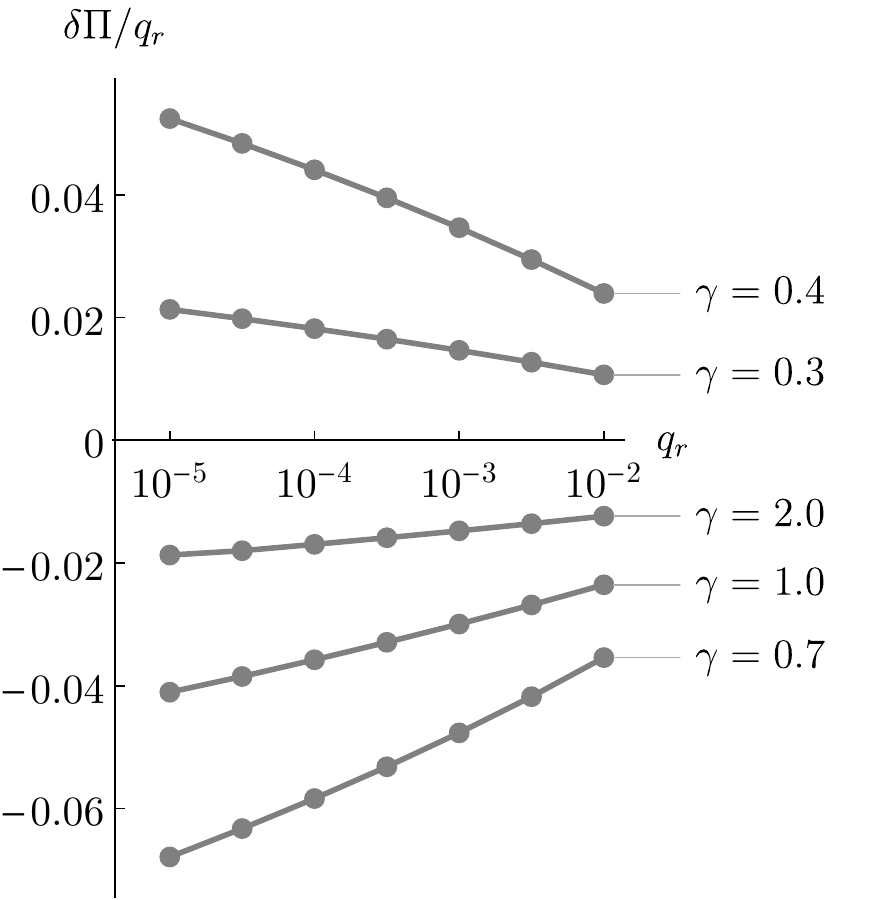}
\caption{Numerical results for $\delta\Pi(\bq,0)/q_r$ computed with a frequency and momentum dependent self-energy for various choices of $\gam = c/b$. The other parameters are $v_F = m = \bar b = 1$, and the multiplicities $M$ and $N$ have also been set equal to one. The energy scale in the logarithms in $\Sg(\bk,0)$ has been chosen as $\Lam = 10^4$, and the integration variables $k_0$, $k_r$, $k_t$ in the polarization function, Eq.~(\ref{Pi}), were restricted by a UV cutoff equal to one. Note the logarithmic scale on the abscissa.}
\end{figure}
To assess the influence of the momentum dependence of the fermion self-energy, we have also computed $\delta\Pi(\bq,0)$ with a momentum-dependent self-energy as obtained in Sec.~III, performing a numerical evaluation of the loop-integral, where we used the perturbative, not resummed, expressions for $\Sg(\bk,0)$. In Fig.~3 we show results for the ratio $\delta\Pi(\bq,0)/q_r$ as a function of $q_r$ for various choices of $\gam = c/b$. One can see that the momentum dependence of $\delta\Pi(\bq,0)$ is essentially linear as in Eq.~(\ref{deltaPi_2/3_a}), and the slope (prefactor of the linear dependence) also diverges logarithmically.
However, in contrast to Eq.~(\ref{deltaPi_2/3_a}), the slope of $\delta\Pi(\bq,0)$ is negative, except for small values of $\gam$. This is simply due to the change of Fermi surface at the hot spots from convex to concave for $\gam > \gam'_c \approx 0.51$, which is imposed by the momentum dependence of $\Sg(\bk,0)$ as discussed in Sec.~III.B.
We recall that Eq.~(\ref{deltaPi_2/3_a}) was derived with a fermion self-energy depending only on frequency. The proportionality of $\delta\Pi(\bq,0)$ to $b$ at fixed $\gam$, as obtained in Eq.~(\ref{deltaPi_2/3_a}), remains true also in the presence of $\Sg(\bk,0)$. The above remarks also hold for $q_r < 0$ (not shown in Fig.~3).

The one-loop result for the susceptibility and the fermion self-energy computed with bare fermion propagators is thus far from being self-consistent. Plugging the one-loop self-energy into the polarization function yields a result that deviates drastically from the bare polarization function $\Pi_0$. This is in striking contrast to the case of the Ising nematic QCP, where the one-loop susceptibility and one-loop fermion self-energy computed with bare propagators are also solutions of the self-consistent one-loop equations for the same quantities computed with full propagators (including the fermion self-energy). \cite{polchinski94,metlitski10_nem}

One might try to find a self-consistent scaling solution of the one-loop equations with a fermion self-energy scaling as $|\om|^\alf$ with an exponent $\alf \neq \frac{2}{3}$. However, we have shown that any momentum-independent non-Fermi liquid self-energy with $\alf < 1$ will spoil the peak of the polarization function at the $2k_F$ wave vector $\bQ$, and hence the mechanism for the density wave instability occurring naturally at a wave vector on the $2k_F$-line. Hence such a scaling solution does not seem to exist.

However, in Sec.~III.B we have seen that the momentum dependence of the fermion self-energy computed with bare polarization bubbles leads to a flattening of the Fermi surface near the hot spots. A comparison with one-dimensional Luttinger liquids \cite{giamarchi} shows that a flat Fermi surface favors a peak in the susceptibilities at $2k_F$. Hence a QCP with a non-Fermi liquid self-energy and a $2k_F$ peak in the susceptibility might be stabilized by a Fermi surface flattening around the hot spots. To describe such a QCP requires a solution for $\Pi(\bq,iq_0)$ and $\Sg(\bk,i\om)$, where momentum and frequency dependences of both quantities are taken into account self-consistently. This is beyond the scope of our present work.
A numerical solution of such self-consistency equations has recently been achieved with rather high momentum and frequency resolution for the case of incommensurate charge order, albeit at low finite temperatures with a small mass term, which regularizes the singularity of the fluctuation propagator. \cite{punk15} There, a small momentum shift of the peak in the susceptibility away from the nesting vector was found, which can be expected for finite temperatures due to the highly asymmetric momentum dependence of the fluctuation propagator. However, it remains unclear whether the shift will persist in the zero temperature limit, that is, at the QCP.

We finally note that vertex corrections do not affect the above results, since there are no singular one-loop vertex corrections for incommensurate density waves.\cite{altshuler95} For ordering wave vectors distinct from half a reciprocal lattice vector, there is no choice of momenta in the vertex correction at which all the propagators are singular. The singular self-energy corrections further suppress the vertex correction at low energies, compared to the one with bare propagators. Ward identities from charge and spin conservation dictate that a singular self-energy implies a singular vertex correction at small momentum transfers, but not at $2k_F$-momenta.  


\section{Conclusion}

In summary, we have analyzed quantum fluctuation effects at the onset of density wave order with an incommensurate $2k_F$ wave vector $\bQ$ in two-dimensional metals -- for the case where $\bQ$ connects only a single pair of hot spots $\pm\bk_H$ on the Fermi surface.
We have confirmed the non-Fermi liquid scaling proportional to $|\om|^{2/3}$ of the imaginary part of the one-loop fermion self-energy $\Sg$ at the hot spots as obtained previously.\cite{holder14}
Computing the momentum and frequency dependence of $\Im\Sg$ also away from the hot spots we obtained the frequency-momentum region governed by non-Fermi liquid behavior. The momentum dependence of the real part $\Re\Sg$ at zero frequency yields a logarithmic divergence of the renormalized Fermi velocity and a logarithmically diverging renormalization of the tangential momentum dependence of the dispersion at the hot spots. Resumming the perturbative divergences by a simple scaling ansatz promotes the logarithms to power-laws, indicating in particular a flattening of the Fermi surface near the hot spots.

Going beyond the leading order perturbation expansion, we have shown that the one-loop result computed with bare fermion propagators is far from being self-consistent. Computing the polarization function with propagators dressed by the one-loop self-energy, one obtains a completely different behavior at the $2k_F$ vector $\bQ$. In particular, the peak at $\bQ$ is destroyed, such that the instability wave vector shifts away from the $2k_F$ line. Moreover, a destruction of the $2k_F$ peak found in the bare polarization function is obtained for any momentum-independent self-energy with a non-Fermi liquid frequency exponent. Hence, a QCP at the onset of a density wave with an incommensurate $2k_F$ wave vector, which is naturally obtained in mean-field theory, seems to be spoiled by fluctuations. However, a $2k_F$ QCP might still be conceivable in a self-consistent solution with a mutual stabilization of a flattened Fermi surface and a $2k_F$ peak in the susceptibility. Exploring this possibility is not easy and is left for future work.

The stability of the $2k_F$ density wave QCP was already discussed by Altshuler {\em et al}. \cite{altshuler95} They found a strong non-renormalizable divergence in the two-loop contributions to the polarization function (computed with bare fermion propagators), and concluded that fluctuations will destroy the QCP in favor of a first order transition. In view of our results we believe that a continuous quantum phase transition might still occur, either at a wave vector shifted away from the $2k_F$ line, or with $2k_F$ fluctuations being enhanced by a flattened Fermi surface.

At the one-loop level, the $2k_F$ density wave QCP is formally equivalent to a QCP separating a normal metal from a Fulde-Ferrell-Larkin-Ovchinikov (FFLO) superconductor, where the one-loop self-energy at the hot spots also scales as $|\om|^{2/3}$ in two dimensions.\cite{piazza16}
Our results thus raise the issue of the stability of the FFLO QCP.
A recent renormalization group analysis indices that the FFLO QCP is stable, at least to leading order in an $\eps$-expansion around $5/2$ dimensions.\cite{pimenov17}
It would be worthwhile to investigate the fate of the FFLO QCP along the lines of our work, too.


\begin{acknowledgments}
We are grateful to Darshan Joshi for a critical reading of the manuscript. T.H.\ is supported by the Minerva Foundation.
\end{acknowledgments}


\begin{appendix}

\section{Effective interaction}

In this appendix we present two concrete examples for the RPA effective interaction, namely for the charge instability in the spinless fermion model with nearest neighbor interactions, and for the spin instability in the Hubbard model.


\subsection{Spinless fermion model}

Spinless lattice fermions with nearest-neighbor interactions exhibit a charge density wave instability, if the interaction $V$ exceeds a certain critical value.
The RPA charge susceptibility reads
\begin{equation}
 \chi(\bq,iq_0) = \frac{\chi_0(\bq,iq_0)}{1 + V(\bq) \chi_0(\bq,iq_0)} \, ,
\end{equation}
where
$\chi_0(\bq,iq_0) = - \Pi_0(\bq,iq_0)$ is the bare charge susceptibility, and
$V(\bq) = 2V (\cos q_1 + \dots + \cos q_d)$ is the Fourier transform of the nearest-neighbor interaction (in $d$ dimensions).
A charge density wave instability occurs at a critical interaction strength $V(\bQ) = -1/\chi_0(\bQ,0)$, where $\bQ$ is the ordering wave vector.
The effective interaction is obtained from the sum of bubble chains as
\begin{equation}
 D(\bq,iq_0) = \frac{V(\bq)}{1 - V(\bq) \Pi_0(\bq,iq_0)} \, .
\end{equation}
At the QCP, this can be approximated by the form~(\ref{D_rpa}) with $g = V(\bQ)$.


\subsection{Hubbard model}

In the Hubbard model, spin-$\frac{1}{2}$ lattice fermions interact via a local interaction $U$ between particles with opposite spin orientation.
The RPA spin susceptibility is given by
\begin{equation}
 \chi(\bq,iq_0) = \frac{\chi_0(\bq,iq_0)}{1 - \frac{1}{2} U \chi_0(\bq,iq_0)} \, ,
\end{equation}
where $\chi_0(\bq,iq_0) = -2 \Pi_0(\bq,iq_0)$. Due to the spin-rotation invariance of the model, it is the same for all spin components.
A spin density wave instability occurs at a critical interaction strength
$U = 2/\chi_0(\bQ,0)$, where $\bQ$ is the ordering wave vector.

\begin{figure}
\centering
\includegraphics[width=8.6cm]{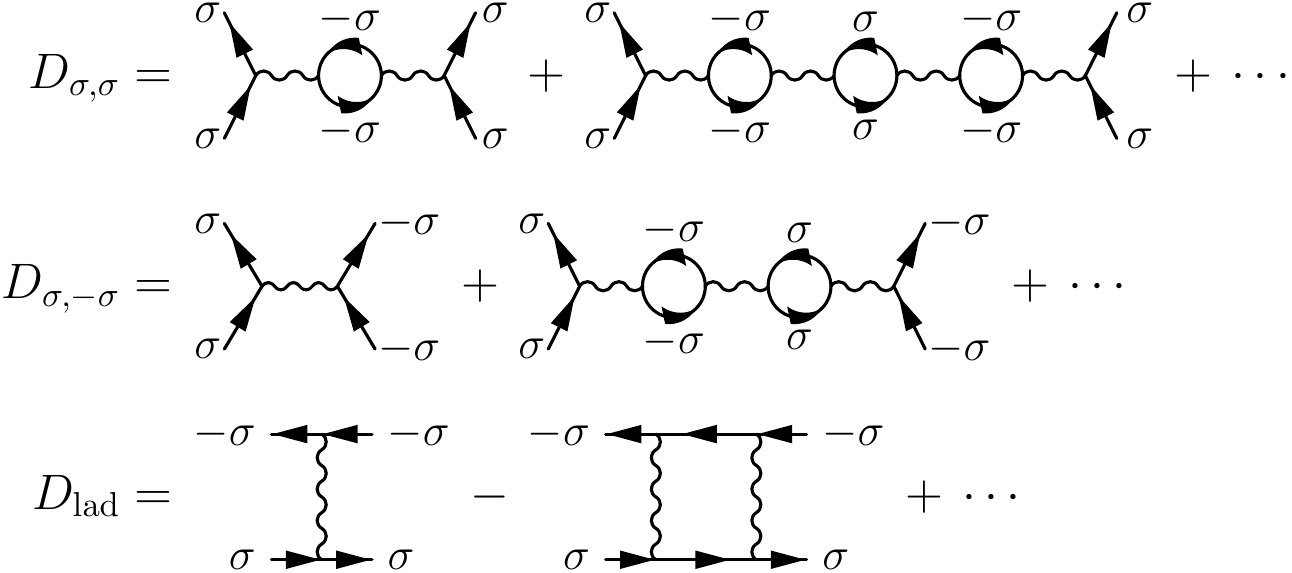}
\caption{Feynman diagrams contributing to the effective interactions in the Hubbard model.}
\end{figure}
Due to the spin structure and the spin rotation invariance, one can construct several effective interactions by summing bubble chains or particle-hole ladders (see Fig.~4), which all diverge at the QCP.
Summing bubble chain diagrams yields an effective interaction between particles with equal spin orientation
\begin{equation}
 D_{\sg,\sg}(\bq,iq_0) =
 \frac{U^2 \Pi_0(\bq,iq_0)}{1 - U^2 [\Pi_0(\bq,iq_0)]^2} \, ,
\end{equation}
and an effective interaction between particles with opposite spin orientation
\begin{equation}
 D_{\sg,-\sg}(\bq,iq_0) =
 \frac{U}{1 - U^2 [\Pi_0(\bq,iq_0)]^2} \, .
\end{equation}
They both diverge at the QCP. Summing particle-hole ladder diagrams, one obtains the effective interaction
\begin{equation}
 D_{\rm lad}(\bq,iq_0) = \frac{U}{1 + U \Pi_0(\bq,iq_0)} \, ,
\end{equation}
which also diverges at the QCP. At the QCP one has $U \Pi_0(\bQ,iq_0) = -1$, such that one can approximate the effective interactions as
\begin{eqnarray}
 D_{\sg,\sg}(\bq,iq_0) &=& D(\bq,iq_0) \, , \\
 D_{\sg,-\sg}(\bq,iq_0) &=& - D(\bq,iq_0) \, , \\
 D_{\rm lad}(\bq,iq_0) &=& -2D(\bq,iq_0) \, ,
\end{eqnarray}
where $D(\bq,iq_0)$ has the form~(\ref{D_rpa}) with $g = -U/2$. The effective interaction can also be written in the manifestly spin-rotation invariant form $D_{\sg'_1\sg'_2\sg_1\sg_2}(\bq,iq_0)$ as in Eq.~(\ref{D_spin}), where the $\tau^z$ component is obtained from the bubble diagrams, while the $\tau^x$ and $\tau^y$ components come from the ladder diagrams. 

\begin{figure}
\centering
\vskip 5mm
\includegraphics[width=9cm]{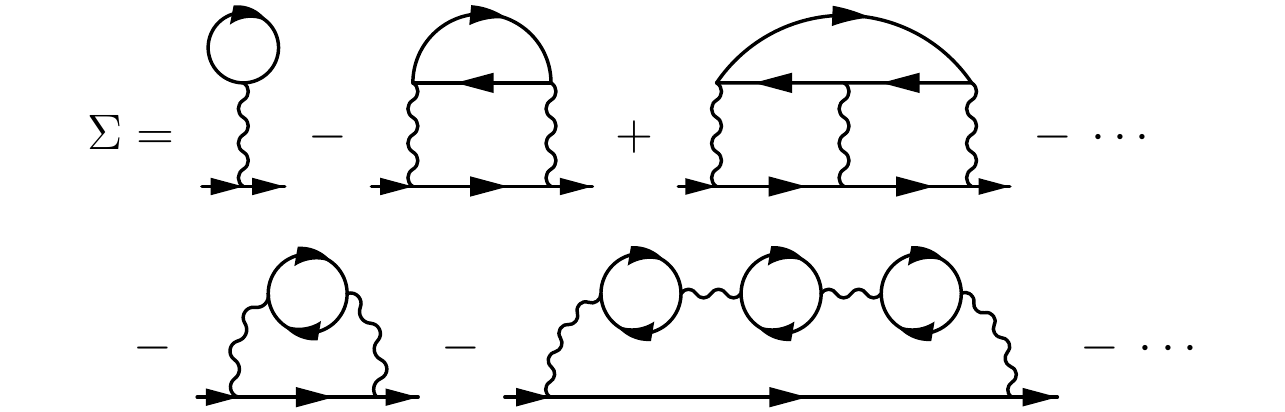}
\caption{Feynman diagrams representing RPA contributions to the fermion self-energy in the Hubbard model.}
\end{figure}
To first order in $D$, there are two distinct contributions to the fermion self-energy where the divergence of $D$ leads to singular contributions (see Fig.~5). They involve $D_{\sg\sg}$ and $D_{\rm lad}$ in the form
\begin{align}
 \Sg(\bk,i\om) =&
 \int \! \frac{d^2\bq}{(2\pi)^2} \int \! \frac{dq_0}{2\pi} \,
 D_{\rm lad}(\bq,iq_0) \, G_0(\bk-\bq,i\om-iq_0) \nonumber \\
 -& \int \! \frac{d^2\bq}{(2\pi)^2} \int \! \frac{dq_0}{2\pi} \,
 D_{\sg,\sg}(\bq,iq_0) \, G_0(\bk-\bq,i\om-iq_0) \, .
 \nonumber \\
\end{align}
Note that in this expression the second order contribution $-\int \frac{d^2\bq}{(2\pi)^2} \int \frac{dq_0}{2\pi} \, U^2 \Pi_0(\bq,iq_0) \, G_0(\bk-\bq,i\om-iq_0)$ appears twice, and should thus be subtracted once. However, keeping only the singular contributions at the QCP, this regular overcounting term can be ignored, and the singular terms can be combined to
\begin{equation}
 \Sg(\bk,i\om) =
 - 3 \int \frac{d^2\bq}{(2\pi)^2} \int \frac{dq_0}{2\pi} \,
 D(\bq,iq_0) \, G_0(\bk-\bq,i\om-iq_0) \, .
\end{equation}
This has the form~(\ref{Sg_oneloop}), with $M=3$.


\begin{widetext}

\section{Computation of one-loop self-energy}

In this appendix we derive the asymptotic results for the fermion self-energy, starting from the one-loop integral Eq.~(\ref{Sg_oneloop}). We set the global multiplicity factor $M$ equal to one for simplicity.
The dispersion relations are expanded as $e_\bq = v_F q_r + \frac{q_t^2}{4m}$ and $\xi_{\bk-\bq} = - v_F (k_r-q_r) + \frac{1}{2m}(k_t-q_t)^2$. Combining these expansions, we can write the fermion energy as
$\xi_{\bk-\bq} = -\xi_\bk + e_\bq + \frac{1}{4m}(q_t - 2k_t)^2$, and use $q_t$ and $e_\bq$ as momentum integration variables.
Inserting the propagators with the above expression for $\xi_{\bk-\bq}$, we obtain the three-fold integral
\begin{equation}
 \Sg(\bk,i\om) = - \frac{1}{av_F}
 \int \frac{de_\bq}{2\pi} \int \frac{dq_t}{2\pi} \int \frac{dq_0}{2\pi} \,
 \frac{1}{-\xi_\bk + e_\bq - i\om + iq_0 + \frac{1}{4m}(q_t - 2k_t)^2} \,
 \frac{1}{h(e_\bq,q_0) - {\bar b} e_\bq + {\bar c} \frac{q_t^2}{4m}} \, ,
\end{equation}
where $h(e_\bq,q_0) = \sqrt{2} \sqrt{\sqrt{e_\bq^2 + q_0^2} + e_\bq}$.
The real part of $\Sg(\bk,i\om)$ requires an ultraviolet regularization (see below).
The $q_t$ integration can be performed by using the residue theorem. The integrand has four simple poles in the complex plane,
\begin{equation}
 q_t = 2k_t \pm 2\sqrt{m} \sqrt{\xi_\bk - e_\bq + i\om - iq_0} \quad \mbox{and} \quad
 q_t = \pm 2i \sqrt{\frac{m}{\bar c}} \sqrt{h(e_\bq,q_0) - {\bar b} e_\bq} \, .
 \nonumber
\end{equation}
Closing the integration contour in the upper or in the lower half-plane yields the same result, but in a different form. To obtain an expression that is manifestly symmetric under $k_t \mapsto -k_t$, we average the two expressions to obtain
\begin{eqnarray} \label{Sg1}
 \Sg(\bk,i\om) &=&
 - \frac{1}{2\pi N} \int de_\bq \int dq_0 \sum_{s = \pm 1}
 \frac{i \, \sgn(\om - q_0)}{\sqrt{\xi_\bk - e_\bq + i\om - iq_0}}
 \frac{1}{h(e_\bq,q_0) - {\bar b} e_\bq
 + {\bar c} \left[ \frac{s k_t}{\sqrt{m}} +
 \sqrt{\xi_\bk - e_\bq + i\om - iq_0} \right]^2}
 \nonumber \\
 && - \frac{1}{2\pi N} \int de_\bq \int dq_0 \sum_{s = \pm 1}
 \frac{1}{\left[ \frac{i}{\sqrt{\bar c}}
 \sqrt{ h(e_\bq,q_0) - {\bar b} e_\bq}
 - \frac{s k_t}{\sqrt{m}} \right]^2 - \xi_\bk + e_\bq - i\om + iq_0} 
 \frac{1}{\sqrt{\bar c} \sqrt{ h(e_\bq,q_0) - {\bar b} e_\bq}} .
 \hskip 7mm
\end{eqnarray}
%


\subsection{Imaginary part}

The dominant contributions to $\Im\Sg$ for small frequencies come from negative $e_\bq$ and $|q_0| \ll |e_\bq|$.\cite{holder14}
We define dimensionless integration variables $\tilde q_0$ and $\tilde e_\bq$ via the relations $q_0 = |\om| \tilde q_0$ and $e_\bq = (|\om|/{\bar b})^{2/3} \tilde e_\bq$, and dimensionless parameters $\tilde\xi_\bk$ and $\tilde k_t$ via $\xi_\bk = (|\om|/{\bar b})^{2/3} \tilde \xi_\bk$ and $k_t = \sqrt{m} (|\om|/{\bar b})^{1/3} \tilde k_t$, respectively.
In the low frequency limit we can approximate $h(\bq,q_0)$ by $|q_0|/\sqrt{|e_\bq|} =
{\bar b}^{1/3} |\om|^{2/3} |\tilde q_0|/\sqrt{|\tilde e_\bq|}$, and
$\xi_\bk - e_\bq + i\om - iq_0 \to \xi_\bk - e_\bq = (|\om|/{\bar b})^{2/3} (\tilde \xi_\bk - \tilde e_\bq)$. The neglected terms are suppressed by a factor of order $|\om|^{1/3}$.
Inserting these approximations with dimensionless variables into Eq.~(\ref{Sg1}), we obtain
\begin{eqnarray}
 \Im\Sg(\bk,i\om) &=& - \frac{1}{2\pi N} \left| \frac{\om}{\bar b} \right|^{2/3}
 \int_{-\infty}^{\min(0,\tilde\xi_\bk)} d{\tilde e_\bq}
 \int_{-\infty}^{\infty} d{\tilde q_0} \,
 \frac{\sgn[\sgn(\om) - \tilde q_0]}{\sqrt{\tilde\xi_\bk - \tilde e_\bq}}
 \nonumber \\
 &\times& \sum_{s = \pm 1} 
 \frac{1}{\displaystyle \frac{|\tilde q_0|}{\sqrt{|\tilde e_\bq|}} - \tilde e_\bq +
 \gam \left[ s \tilde k_t + \sqrt{\tilde\xi_\bk - \tilde e_\bq} \right]^2} \, ,
\end{eqnarray}
with $\gam = \bar c/\bar b = c/b$.
The integration over $\tilde q_0$ is now elementary. Using
\begin{equation}
 \int_{-\infty}^{\infty} d\tilde q_0 \,
 \frac{\sgn(\pm 1 - q_0)}{A|\tilde q_0| + B} =
 \pm 2 \int_0^1 \frac{d\tilde q_0}{A \tilde q_0 + B} =
 \pm \frac{2}{A} \ln(1 + A/B) \, ,
\end{equation}
we obtain
\begin{equation} \label{A:ImSg}
 \Im\Sg(\bk,i\om) =
 - \frac{\sgn(\om)}{\pi N} \left| \frac{\om}{\bar b} \right|^{2/3}
 I(\tilde\xi_\bk,\tilde k_t) \, ,
\end{equation}
with the dimensionless scaling function
\begin{equation}
 I(\tilde\xi_\bk,\tilde k_t) =
 \int_{-\infty}^{\min(0,\tilde\xi_\bk)}
 \frac{d\tilde e_\bq}{\sqrt{1 - \tilde\xi_\bk/\tilde e_\bq}}
 \sum_{s = \pm 1} \ln \left[ 1 +
 \frac{|\tilde e_\bq|^{-3/2}}
 {1 + \gam \Big( s \, \tilde k_t/\sqrt{|\tilde e_\bq|} +
 \sqrt{1 - \tilde\xi_\bk/\tilde e_\bq} \, \Big)^2} \right] \, .
\end{equation}

We now evaluate the scaling function analytically in various limiting cases.
For $\bk = \bk_H$ one has $\tilde\xi_\bk = 0$ and $\tilde k_t = 0$. From
\begin{equation}
 I(0,0) = 2 \int_{-\infty}^0 d \tilde e_\bq \, 
 \ln \left( 1 + \frac{|\tilde e_\bq|^{-3/2}}{1 + \gam} \right) =
 \frac{4\pi}{\sqrt{3}} \frac{1}{(1 + \gam)^{2/3}}
\end{equation}
one obtains Eq.~(\ref{ImSg_kH}) for $\Im\Sg(\bk_H,i\om)$. The same result is valid for
$\bar b |\xi_\bk|^{3/2} \ll |\om|$ and $\bar b \frac{|k_t|^3}{m^{3/2}} \ll |\om|$, because under these conditions one can approximate $\tilde\xi_\bk$ and $\tilde k_t$ by zero.

For $k_t = 0$ and $\xi_\bk < 0$ one has $\tilde k_t = 0$, $\tilde\xi_\bk < 0$, and thus
\begin{equation}
 I(\tilde\xi_\bk,0) = 2 \int_{-\infty}^{\tilde\xi_\bk}
 \frac{d\tilde e_\bq}{\sqrt{1 - \tilde\xi_\bk/\tilde e_\bq}}
 \ln \left[ 1 + \frac{|\tilde e_\bq|^{-3/2}}
 {1 + \gam (1 - \tilde\xi_\bk/\tilde e_\bq)}
 \right] \, .
\end{equation}
Substituting $x = \tilde e_\bq/\tilde\xi_\bk$, and expanding for $|\tilde\xi_\bk| \gg 1$, one obtains
\begin{eqnarray}
 I(\tilde\xi_\bk,0) &=&
 2 |\tilde\xi_\bk| \int_1^\infty \frac{dx}{\sqrt{1 - x^{-1}}} \,
 \ln \left[ 1 + |\tilde\xi_\bk|^{-3/2} \frac{x^{-1/2}}{(1+\gam)x - \gam} \right]
 \nonumber \\[2mm]
 &\approx& \frac{2}{\sqrt{|\tilde\xi_\bk|}} \int_1^\infty
 \frac{dx}{\sqrt{x-1}} \, \frac{1}{(1+\gam) x - \gam} =
 \frac{2\pi}{\sqrt{1 + \gam}} \frac{1}{\sqrt{|\tilde\xi_\bk|}} \, .
\end{eqnarray}
Inserting this into Eq.~(\ref{A:ImSg}) yields Eq.~(\ref{ImSg_xik-}).
For $k_t = 0$ and $\xi_\bk > 0$ one has $\tilde k_t = 0$, $\tilde\xi_\bk > 0$, and thus
\begin{equation}
 I(\tilde\xi_\bk,0) = 2 \int_{-\infty}^0
 \frac{d\tilde e_\bq}{\sqrt{1 - \tilde\xi_\bk/\tilde e_\bq}}
 \ln \left[ 1 + \frac{|\tilde e_\bq|^{-3/2}}
 {1 + \gam (1 - \tilde\xi_\bk/\tilde e_\bq)}
 \right] \, .
\end{equation}
Substituting $x = |\tilde e_\bq|/\tilde\xi_\bk$ and expanding for $\tilde\xi_\bk \gg 1$ and $\tilde\xi_\bk^{3/2} \gg \gam^{-1}$, one obtains
\begin{eqnarray}
 I(\tilde\xi_\bk,0) &=&
 2 \xi_\bk \int_0^\infty \frac{dx}{\sqrt{1 + x^{-1}}} \,
 \ln \left[ 1 + \xi_\bk^{-3/2} \frac{x^{-1/2}}{(1+\gam)x + \gam} \right]
 \nonumber \\[2mm]
 &\approx& \frac{2}{\sqrt{\tilde\xi_\bk}} \int_0^\infty
 \frac{dx}{\sqrt{x+1}} \, \frac{1}{(1 + \gam) x + \gam} =
 \frac{4\mbox{Arcsinh}(\gam^{-1/2})}{\sqrt{1 + \gam}}
 \frac{1}{\sqrt{\tilde\xi_\bk}} \, .
\end{eqnarray}
Inserting this into Eq.~(\ref{A:ImSg}) yields Eq.~(\ref{ImSg_xik+}). Note that the condition $\tilde\xi_\bk \gg 1$ alone is not sufficient to justify the above expansion. 
Equations~(\ref{ImSg_xik-}) and (\ref{ImSg_xik+}) hold also for $k_t \neq 0$ with
$\bar b \frac{|k_t|^3}{m^{3/2}} \ll |\om|$, because under this condition one can approximate $\tilde k_t$ by zero.

For $\xi_\bk = 0$ and $k_t \neq 0$ we have $\tilde\xi_\bk = 0$ and $\tilde k_t \neq 0$, and thus
\begin{equation}
 I(0,\tilde k_t) = \int_{-\infty}^0 d \tilde e_\bq \sum_{s = \pm 1}
 \ln \left[ 1 + \frac{|\tilde e_\bq|^{-3/2}}{1 +
 \gam \left( 1 + s \tilde k_t/\sqrt{|\tilde e_\bq|} \right)^2} \right] \, .
\end{equation}
Substituting $x = |\tilde e_\bq|/\tilde k_t^2$, and expanding for $|\tilde k_t| \gg 1$ and $|\tilde k_t|^3 \gg \gam^{-1}$, one obtains
\begin{eqnarray}
 I(0,\tilde k_t) &=& \tilde k_t^2 \int_0^\infty dx \sum_{s = \pm 1}
 \ln \left[ 1 + \frac{1}{|\tilde k_t|^3} \, 
 \frac{x^{-1/2}}{(1+\gam)x + \gam + 2s\gam\sqrt{x}} \right]
 \nonumber \\[2mm]
 &\approx& \frac{1}{|\tilde k_t|} \int_0^\infty \frac{dx}{\sqrt{x}}
 \sum_{s = \pm 1} \frac{1}{(1+\gam)x + \gam + 2s\gam\sqrt{x}} =
 \frac{2\pi}{\sqrt{\gam}} \, \frac{1}{|k_t|} \, . 
\end{eqnarray}
Inserting this into Eq.~(\ref{A:ImSg}) yields Eq.~(\ref{ImSg_kt}). The result remains valid for $\xi_\bk \neq 0$ as long as $\bar b |\xi_\bk|^{3/2} \ll |\om|$ since then one can approximate $\tilde\xi_\bk$ by zero.


\subsection{Real part}

We now compute the momentum dependence perpendicular to the Fermi surface of the self-energy at zero frequency. Setting $\om = 0$ and $k_t = 0$ in Eq.~(\ref{Sg1}), one obtains
\begin{equation}
 \Sg(\bk,0) = \frac{1}{\pi N} \int de_\bq \int dq_0 \,
 \frac{1}{h(e_\bq,q_0) - \bar b e_\bq + \bar c (\xi_\bk - e_\bq - iq_0)} 
 \left[ \frac{i\sgn(q_0)}{\sqrt{\xi_\bk - e_\bq - iq_0}} +
 \frac{\sqrt{\bar c}}{\sqrt{h(e_\bq,q_0) - \bar b e_\bq}} \right] \, .
\end{equation}
This is a sum of two (real) terms, which we denote by $\Sg_1(\bk,0)$ and $\Sg_2(\bk,0)$.
For small $\xi_\bk$, the leading contributions of both terms to $\delta\Sg(\bk,0) =
\Sg(\bk,0) - \Sg(\bk_H,0)$ are proportional to $|\xi_\bk| \ln|\xi_\bk|$.

Imposing an ultraviolet cutoff $\sqrt{e_\bq^2 + q_0^2} \leq \Lam$ and introducing rescaled integration variables defined by $e_\bq = |\xi_\bk| \tilde e_\bq$ and $q_0 = |\xi_\bk| \tilde q_0$, the first term can be written as
\begin{eqnarray}
 \Sg_1(\bk,0) &=& \frac{|\xi_\bk|}{\pi N} 
 \int d\tilde e_\bq \int d\tilde q_0 \,
 \Theta\left(\Lam/|\xi_\bk| - \sqrt{\tilde e_\bq^2 + \tilde q_0^2} \right) \,
 \frac{i \sgn(\tilde q_0)}{\sqrt{\sgn(\xi_\bk) - \tilde e_\bq - i \tilde q_0}} 
 \nonumber \\
 &\times& 
 \frac{1}{\sqrt{2} \sqrt{\sqrt{\tilde e_\bq^2 + \tilde q_0^2} + \tilde e_\bq}
  + \sqrt{|\xi_\bk|} \left[ \bar c (\sgn(\xi_\bk) - \tilde e_\bq - i\tilde q_0)
  - \bar b \tilde e_\bq \right]} \, .
\end{eqnarray}
Dominant contributions for $\xi_\bk \to 0$ come from large (UV) and small (IR) integration variables $\tilde q_0$ and $\tilde e_\bq$. To compute the UV contribution, it is convenient to use polar coordinates $r$ and $\theta$ such that
$\tilde e_\bq + i\tilde q_0 = r e^{i\theta}$.
Subtracting $\Sg_1(\bk_H,0)$ and letting $\xi_\bk \to 0$ in the integrand, one obtains
\begin{equation}
 \delta\Sg_1^{\rm UV}(\bk,0) = \frac{|\xi_\bk|}{\pi N}
 \int_0^{\Lam/|\xi_\bk|} dr \int_{-\pi}^{\pi} d\theta \,
 \frac{i \sqrt{r} \, \sgn(\theta)}{2\cos(\theta/2)} \left(
 \frac{1}{\sqrt{\sgn(\xi_\bk) - r e^{i\theta}}} - \frac{1}{\sqrt{-r e^{i\theta}}}
 \right) \, .
\end{equation}
Note that $\delta\Sg_1^{\rm UV}(\bk,0)$ does not depend on $\bar b$ and $\bar c$.
For large $r$, the integrand decays as $r^{-1}$, which yields a logarithmic UV divergence. The prefactor is determined by an elementary integration over $\theta$, yielding
\begin{equation}
 \delta\Sg_1^{\rm UV}(\bk,0) =
 \frac{1}{2N} \, \xi_\bk \ln\frac{\Lam}{|\xi_\bk|} +
 {\cal O}(\xi_\bk) \, .
\end{equation}
For $\xi_\bk < 0$ there is also a contribution of the same order from
$-1 \leq \tilde e_\bq \leq 0$ and $|\tilde q_0| \ll |\tilde e_\bq|$.
Expanding $\sqrt{2} \sqrt{\sqrt{\tilde e_\bq^2 + \tilde q_0^2} + \tilde e_\bq} \approx
|\tilde q_0|/\sqrt{|\tilde e_\bq|}$, this contribution can be written as
\begin{equation}
 \Sg_1^{\rm IR}(\bk,0) =
 - \frac{|\xi_\bk|}{\pi N} \int_{-1}^0 d\tilde e_\bq
 \int_{-\lam |\tilde e_\bq|}^{\lam |\tilde e_\bq|} d\tilde q_0
 \frac{1}{\sqrt{1 + \tilde e_\bq}} \, \frac{1}{|\tilde q_0|/\sqrt{\tilde e_\bq} - 
 \sqrt{|\xi_\bk|} [\bar c(1 + \tilde e_\bq) + \bar b \tilde e_\bq]}
\end{equation}
for small $\xi_\bk$, where $\lam$ is a small positive number.
The frequency integration is now elementary, leading to
\begin{equation}
 \Sg_1^{\rm IR}(\bk,0) = - \frac{|\xi_\bk|}{\pi N} \int_{-1}^0 d\tilde e_\bq
 \frac{2\sqrt{|\tilde e_\bq|}}{\sqrt{1 + \tilde e_\bq}}
 \ln\left| 1 - \frac{\lam \sqrt{|\tilde e_\bq|}}{\sqrt{|\xi_\bk|}
 [\bar c(1 + \tilde e_\bq) + \bar b \tilde e_\bq]} \right| \, .
\end{equation}
Using
$\int_{-1}^0 d\tilde e_\bq \, \frac{\sqrt{|\tilde e_\bq|}}{\sqrt{1 + \tilde e_\bq}} =
 \pi/2$,
the leading contribution for $\xi_\bk \to 0$ can be explicitly evaluated, yielding
\begin{equation}
 \Sg_1^{\rm IR}(\bk,0) = - \frac{|\xi_\bk|}{2N} \ln \frac{\Lam}{|\xi_\bk|} +
 {\cal O}(\xi_\bk) \, .
\end{equation}
The choice of the arbitrary constants $\Lam$ and $\lam$ affects only the subleading term of order $\xi_\bk$. We recall that $\Sg_1^{\rm IR}(\bk,0)$ contributes only for $\xi_\bk < 0$. Hence,
\begin{equation}
 \delta\Sg_1(\bk,0) = 
 \frac{1}{2N} \left[ 1 + \Theta(-\xi_\bk) \right] \xi_\bk \ln \frac{\Lam}{|\xi_\bk|} +
 {\cal O}(\xi_\bk) \, .
\end{equation}

Making the substitution $\sqrt{e_\bq + iq_0} = x + iy$, the second contribution to $\Sg(\bk,0)$ can be written as
\begin{equation}
 \Sg_2(\bk,0) = \frac{1}{\pi N} \int_0^{\sqrt{\Lam}} dx \int_{-\infty}^{\infty} dy \,
 \frac{4(x^2+y^2)}{\sqrt{2x - \bar b (x^2-y^2)}}
 \frac{1}{2x - \bar b (x^2-y^2) + \bar c [\xi_\bk - (x+iy)^2]} \, ,
\end{equation}
where $\Lam$ is an ultraviolet cutoff with dimension of energy.
The imaginary part in the denominator does not contribute to leading order and can thus be ignored. Subtracting $\Sg_2(\bk_H,0)$ and performing the $y$-integration yields
\begin{equation}
 \delta\Sg_2(\bk,0) = - \frac{8}{\pi N (\bar b + \bar c)} \int_0^{\sqrt{\Lam}} dx
 \left[ g \Big( 
 \frac{2x - 2(\bar b + \bar c)x^2 + \bar c \xi_\bk}{2x - \bar b \xi_\bk} \Big) -
 g \left( 1 - (\bar b + \bar c) x \right) \right] \, ,
\end{equation}
where $g(u) = \sqrt{u} \arctan \sqrt{\bar c/(\bar b u)}$.
Introducing the dimensionless variable $\tilde x$ defined by
$x = \bar b |\xi_\bk| \tilde x$, this can be written as
\begin{equation}
 \delta\Sg_2(\bk,0) = - \frac{8 |\xi_\bk|}{\pi N (1+\gam)} 
 \int_0^{\frac{\sqrt{\Lam}}{\bar b |\xi_\bk|}} d \tilde x \left[
 g \Big( \frac{2 \tilde x - 2(1+\gam)\bar b^2 |\xi_\bk| \tilde x^2 + \gam \, \sgn\xi_\bk}
 {2 \tilde x - \sgn\xi_\bk} \Big) -
 g \left( 1 - (1+\gam) \bar b^2 |\xi_\bk| \tilde x \right) \right] ,
\end{equation}
with $\gam = \bar c/\bar b$.
The leading contributions to this integral come from large $\tilde x$. Restricting the integration to $\tilde x \geq 1$, and expanding for large $\tilde x$ and small $\xi_\bk$, one obtains
\begin{equation}
 \delta\Sg_2(\bk,0) = - \frac{8 |\xi_\bk|}{\pi N (1+\gam)} 
 \int_1^{\frac{\sqrt{\Lam}}{\bar b |\xi_\bk|}} d \tilde x \,
 g'(1) \, \frac{(1+\gam) \, \sgn\xi_\bk}{2 \tilde x}
 = \frac{2}{\pi N} \Big( \frac{\sqrt{\gam}}{1+\gam} - \arctan\sqrt{\gam} \Big)
 \xi_\bk \ln\frac{\sqrt{\Lam}}{\bar b |\xi_\bk|} \, .
\end{equation}
Summing $\delta\Sg_1(\bk,0)$ and $\delta\Sg_2(\bk,0)$ yields Eq.~(\ref{ReSg_xik}).


\section{Computation of polarization bubble}

In the following we evaluate the polarization bubble $\Pi(\bq,iq_0)$ defined in Eq.~(\ref{Pi}) with a self-energy of the form (\ref{Sg_alf}).
Using radial and tangential momentum coordinates and expanding the fermion dispersion yields 
\begin{equation}
 \Pi(\bq,iq_0) = \int \frac{d^2\bk}{(2\pi)^2} \int \frac{dk_0}{2\pi} \,
 \frac{1}{i\{k_0\} - v_F k_r - \frac{k_t^2}{2m}} \,
 \frac{1}{i\{k_0 - q_0\} + v_F(k_r - q_r) - \frac{(k_t - q_t)^2}{2m}} \, .
\end{equation}
Here we have introduced the short-hand notation $\{\om\} = C \sgn(\om) |\om|^\alf$.
The $k_r$ and $k_t$ integrations can be performed by using the residue theorem,
\begin{eqnarray}
 \Pi(\bq,iq_0) &=& \frac{i}{v_F} \int \frac{dk_0}{2\pi} \frac{dk_t}{2\pi} \,
 \frac{\Theta(-k_0) - \Theta(k_0 - q_0)}
 {i(\{k_0\} + \{k_0 - q_0\}) - v_F q_r - \frac{k_t^2}{2m} - \frac{(k_t - q_t)^2}{2m}}
 \nonumber \\[2mm]
 &=& \frac{i}{v_F} \int \frac{dk_0}{2\pi} \int \frac{dk_t}{2\pi} \,
 \frac{\Theta(k_0 - q_0) - \Theta(-k_0)}
 {\frac{(k_t - q_t/2)^2}{m} + e_\bq - i(\{k_0\} + \{k_0 - q_0\})}
 \nonumber \\[2mm]
 &=& - \frac{\sqrt{m}}{4\pi v_F} \int dk_0 \,
 \frac{|\Theta(k_0 - q_0) - \Theta(-k_0)|}
 {\sqrt{-e_\bq + i(\{k_0\} + \{k_0 - q_0\})}} \, .
\end{eqnarray}
Subtracting $\Pi(\bQ,0)$ one thus obtains
\begin{equation} \label{deltaPi_dyn}
 \delta\Pi(\bq,iq_0) = - \frac{\sqrt{m}}{4\pi v_F} \left[
 \int dk_0 \, \frac{|\Theta(k_0 - q_0) - \Theta(-k_0)|}
 {\sqrt{-e_\bq + i(\{k_0\} + \{k_0 - q_0\})}} -
 \int \frac{dk_0}{\sqrt{2i\{k_0\}}} \right] \, .
\end{equation}

For the bare bubble one obtains the same expression with $\{k_0\} = k_0$. The $k_0$-integration is then convergent (without UV cutoff) and elementary, yielding the familiar singular contribution
\begin{equation}
 \delta\Pi_0(\bq,iq_0) = \frac{\sqrt{m}}{4\pi v_F}
 \left( \sqrt{e_\bq + iq_0} + \sqrt{e_\bq - iq_0} \right) \, .
\end{equation}

We now evaluate Eq.~(\ref{deltaPi_dyn}) for $\alf < 1$ in the static limit $q_0 = 0$. Introducing an ultraviolet cutoff $\Lam$, we can write $\delta\Pi(\bq,0)$ as
\begin{equation}
 \delta\Pi(\bq,0) = - \frac{\sqrt{m}}{2\pi v_F} \Re \int_0^\Lam dk_0 \left(
 \frac{1}{\sqrt{-e_\bq + 2i\{k_0\}}} - \frac{1}{\sqrt{2i\{k_0\}}}
 \right) \, .
\end{equation}
Using dimensionless variables $\tilde k_0$ and $\tilde\Lam$ defined via
\begin{equation}
 k_0 = |C^{-1} e_\bq \tilde k_0|^{1/\alf} \, ,
 \Lam = |C^{-1} e_\bq \tilde\Lam|^{1/\alf} \, ,
\end{equation}
we can scale out dimensionful quantities as
\begin{equation} \label{deltaPi_3}
 \delta\Pi(\bq,0) =
 \frac{\sqrt{m}|e_\bq|^{\frac{1}{\alf} - \frac{1}{2}}}
 {2\pi v_F C^{\frac{1}{\alf}}} \delta\tilde\Pi(\bq,0) \, ,
\end{equation}
where
\begin{eqnarray} \label{Pitilde}
 \delta\tilde\Pi(\bq,0) &=&
 - \frac{1}{\alf} \, \Re \int_0^{\tilde\Lam} d\tilde k_0 \,
 \tilde k_0^{\frac{1}{\alf} - 1} \left(
 \frac{1}{\sqrt{2i \tilde k_0 - \sgn(e_\bq)}} - \frac{1}{\sqrt{2i \tilde k_0}}
 \right) \nonumber \\
 &=& - \Re \, \frac{1}{\alf \sqrt{2i}} \int_0^{\tilde\Lam} d\tilde k_0 \,
 \tilde k_0^{\frac{1}{\alf} - \frac{3}{2}} \left(
 \frac{1}{\sqrt{1 - \frac{\sgn(e_\bq)}{2i \tilde k_0}}} - 1 \right) \, .
\end{eqnarray}

For $\alf > \frac{2}{3}$ the integral is UV convergent, that is, we can set $\tilde\Lam \to \infty$. From
\begin{equation}
 \int_0^\infty d\tilde k_0 \, \tilde k_0^{\frac{1}{\alf} - \frac{3}{2}}
 \left( \frac{1}{\sqrt{1 \mp \frac{1}{2i \tilde k_0}}} - 1 \right) =
 \frac{1}{\sqrt{\pi}} \left(\pm \frac{i}{2} \right)^{\frac{1}{\alf} - \frac{1}{2}}
 \Gamma \Big( \frac{1}{2} - \frac{1}{\alf} \Big) \,
 \Gam \Big( \frac{1}{\alf} \Big) \, ,
\end{equation}
where $\Gamma$ is the Gamma function, one thus obtains Eq.~(\ref{deltaPi_alf}), with the coefficients
\begin{equation} \label{apm}
 a_\pm = \frac{\sqrt{m}}{2\pi v_F}
 \frac{2^{-\frac{1}{\alf}}}{\sqrt{\pi} \alf C^{\frac{1}{\alf}}} \,
 \cos \Big[ \frac{\pi}{4} \Big( 3 \mp 1 \pm \frac{2}{\alf} \Big) \Big] \,
 \Gamma \Big( \frac{1}{2} - \frac{1}{\alf} \Big) \,
 \Gam \Big( \frac{1}{\alf} \Big) \, .
\end{equation}
Note that $a_+ \to \sqrt{m}/(2\pi v_F C)$ and $a_- \to 0$ for $\alf \to 1$.

For $\alf = \frac{2}{3}$, the integral in Eq.~(\ref{Pitilde}) is dominated by large $\tilde k_0$, so that we can expand
\begin{eqnarray}
 \delta\tilde\Pi(\bq,0) &=&
 - \Re \, \frac{3}{2 \sqrt{2i}} \int_0^{\tilde\Lam} d\tilde k_0 \,
 \left( \frac{1}{\sqrt{1 - \frac{\sgn(e_\bq)}{2i \tilde k_0}}} - 1 \right)
 \nonumber \\
 &\approx& - \Re \, \frac{3}{2 \sqrt{2i}}
 \int_1^{\tilde\Lam} d\tilde k_0 \,
 \frac{\sgn(e_\bq)}{4i \tilde k_0} =
 \frac{3}{16} \, \sgn(e_\bq) \ln(\tilde\Lam) \, .
\end{eqnarray}
Inserting this into Eq.~(\ref{deltaPi_3}) with $\alf = \frac{2}{3}$ yields Eq.~(\ref{deltaPi_2/3}).

\end{widetext}

\end{appendix}



\begin{thebibliography}{99}
\bibitem{footnote1} Commensurate wave vectors can be written as a linear
 combination of reciprocal lattice vectors with rational coefficients,
 while incommensurate wave vectors cannot. Obviously there are infinitly
 many possible rational coefficients, but the quantitative effects of
 commensurability decrease with the size of their denominators.
\end{thebibliography}
\end{document}